\documentclass[twocolumn,letterpaper,showpacs,pre,superscriptaddress]{revtex4}
\usepackage{graphicx}
\usepackage{amstext}
\usepackage{amsbsy}
\usepackage{amssymb}
\usepackage{algorithmic}
\usepackage{algorithm}

\begin{document}

\title{Steady-State Properties of a Totally Asymmetric Exclusion Process with 
Periodic Structure}

\author{Greg Lakatos}
\email{glakatos@physics.ubc.ca}
\affiliation{Department of Physics, University of British Columbia, Vancouver, BC, Canada, V6T-1Z1}

\author{Tom Chou}
\email{tomchou@ucla.edu}
\affiliation{Department of Biomathematics, UCLA, Los Angeles, CA 90095-1766}

\author{Anatoly Kolomeisky}
\email{tolya@rice.edu}
\affiliation{Department of Chemistry, Rice University, Houston, TX  77251-1892}

\date{\today}

\begin{abstract}

We formulate and analyze the steady-state behavior of totally asymmetric simple exclusion
processes (TASEPs) that contain periodically varying movement rates.  In our models,
particles at a majority sites hop to the right with rate $p_1$ while particles occupying 
a periodically arranged set of sites move to the right at rate $p_2$.
A number of approximate mean field approaches
are used to study the steady-state currents and bulk densities of this model.  While exact
solutions are not found, the mean field approaches provide results that show good
agreement with data derived from extensive Monte-Carlo simulations.

\end{abstract}


\maketitle

\section{Introduction}

As a canonical model of one-dimensional transport, the Totally Asymmetric Simple
Exclusion Process (TASEP) is a topic of current interest.  One of the few
solvable non-equilibrium models, the TASEP and its solutions have been
extensively investigated \cite{DER92,DER93,DOMANY,DER98,KOLO} as a model
for numerous one-dimensional transport processes including ribosome motion
\cite{SHAW, MACDONALD, LAKATOS}, pore transport
\cite{LOHSE}, and traffic flow \cite{SCHRECKENBERG,NAGATANI}.  The traditional TASEP model
consists of a finite lattice with open boundaries.  Particles are inserted into an empty site
at the leftmost end of the lattice at a rate $\alpha$, and once in the lattice the particles
move to the right at a rate $p$.  Each motion of a particle within the lattice moves
that particle exactly one lattice site to the right, and particles can move only if
the site immediately to the right is not occupied by another particle.  Upon reaching the
rightmost lattice site, particles are removed from the lattice at a rate $\beta$.  Exact
solutions for the steady-state particle currents $\left(J\right)$, and densities
$\left(\sigma\right)$, are available through mean field approaches
\cite{DER92,MACDONALD}, matrix product methods \cite{DER93}, and
recurrence relations \cite{DOMANY}.  The latter two approaches solve the model
completely, providing exact expressions for density correlation functions of
arbitrary order.  The TASEP solution itself yields a phase diagram with three
phases.  At large values of the injection $\left(\alpha\right)$ and extraction
$\left(\beta\right)$ rates the system is dominated by the rate of particle hopping
and is in a maximal current phase.  At small values of $\alpha$ and $\beta$, the
system is found in a low-density entry limited phase, and a high-density
exit-limited phase respectively.  Within each of these two boundary-limited phases
lie a pair of sub-phases each with a distinct density profile.  The fundamental form
of this phase diagram has proven to be very robust and extending the TASEP to
include particles that occlude more than one lattice site \cite{LAKATOS}, or backwards particle motions
\cite{DER98}, has not altered the phase diagram significantly.  Nonetheless, these extensions have
facilitated the use of the TASEP in modeling a wide variety of physical processes.
In many phenomena of interest however, the assumption of a single internal
hopping rate $p$ implicit in the normal TASEP does not adequately capture the full
character of the transport behavior.  
One system that has been the subject of intense interest is the kinesin motor
protein and related molecules \cite{BLOCK}.  Kinesin, and kinesin-like molecules, form an integral
part of the cellular transport system, shuttling protein-filled lipid vesicles to
various parts of the cell.  Traveling along tubulin polymers, the motion of the 
kinesin molecule is driven by a complex conformational change producing steps
approximately $8$nm in length \cite{SCHLIWA}.  The exact stepping mechanism of
the kinesin family of motor proteins is currently the subject of speculation,
however one of the leading candidate models is the so-called "inchworm" model \cite{HUA}.  The inchworm
model proposes that a single $8$nm step is composed of two sub-steps; in the first sub-step
one kinesin head group moves from its current position to the next available binding site on
the tubulin filament. This sub-step is immediately followed by a dragging motion
where the lagging head group is
pulled up to essentially the same position as the lead group.  

\begin{figure}[h]
  \begin{center}
    \includegraphics[width=3.0in]{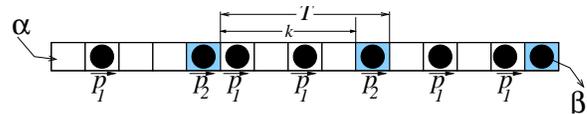}
  \end{center}
  \caption{The two-rate totally asymmetric exclusion model.  At every $T$ lattice sites, the
   particle movement rate is $p_2$.  At all intervening sites, the movement rate is $p_1$.
   All other aspects of the model are identical to those of the standard TASEP.}
  \label{DUALRATE}
\end{figure}

While TASEP models have
been successfully applied to model molecular motor dynamics in the past \cite{PARMEGGIANI}, applying the
TASEP to the two-part motion of the inchworm model requires that the TASEP be extended.
As the two sub-steps in the inchworm model are not expected
to occur at the same rate we can expand the applicability of the TASEP to these types of systems 
by introducing two internal hopping rates $p_1$,
and $p_2$ (cf. Fig. \ref{DUALRATE}) into the TASEP model.  We assign these hopping rates to the lattice sites in
an alternating fashion where odd indexed lattice sites move particles to the right
with rate $p_1$, while even indexed sites move particles with rate $p_2$.  Using this arrangement
a single $(p_1,p_2)$ unit would comprise, for example, a $8$nm kinesin step with the $p_1$ step representing
the movement of the lead head group, and the $p_2$ step the movement of the trailing group.
Generalizing this two-rate TASEP model, we also consider the case where the $p_2$ sites are
separated by an arbitrary number ($k$) of $p_1$ sites.
Applying mean field methods we develop accurate approximations to the
steady-state current and density of the dual-rate TASEP in the maximal current phase,
entry limited, and exit limited regimes.  We also conjecture that, as was true with
many previous extensions to the standard TASEP, the dual-rate extension does
not fundamentally alter the form the TASEP phase diagram.  The accuracy of our 
mean field methods are verified through
extensive Monte Carlo simulations.

\section{Mean Field Theories}

\subsection{Simple Mean Field Methods}

We begin by considering the dual-rate TASEP 
in the limit of large $\alpha$ and $\beta$.  As is the case in the
standard TASEP, the dynamics of the system will be determined entirely by the
internal movement rates $p_1$ and $p_2$.  
Ensuring continuity of the current in a lattice with $k$ $p_1$ sites between
each $p_2$ site we find

\begin{equation}
  J = p_2 \sigma_{k+1} (1 - \sigma_1) = p_1 \sigma_k (1 - \sigma_{k+1}) = \ldots
  \label{SMFT1}
\end{equation}

In writing Eqn (\ref{SMFT1}) we have assumed that the densities ($\sigma_i$) in the lattice
have the same period ($T \equiv (k + 1)$) as the movement rates in the lattice 
\footnote{If the instantaneous lattice occupancy is $\tau_i \in [0,1]$, the average density at site $i$
in the steady state is defined as $\sigma_i = <\tau_i>$}.
Moreover, as the two-rate TASEP preserves the particle-hole symmetry found in the standard TASEP, we expect
for a lattice of size $N$ that $\sigma_i = (1 - \sigma_{N - i + 1})$ \cite{KOLO}.  Coupling this condition
to the assumption of periodicity in the densities, we find that within a single period,
$\sigma_i = (1 - \sigma_{k - i + 2})$.  Thus we can immediately eliminate half the variables in 
(\ref{SMFT1}).  While straight-forward to solve, (\ref{SMFT1}) yields increasingly
unwieldy expressions for the current and the densities as $k$ increases.
As a result we only show the result for the $k = 1$ case:

\begin{equation}
  \begin{array}{rl}
	  J & \displaystyle =  \frac{p_1 p_2}{\left(\sqrt{p_1} + \sqrt{p_2}\right)^2} \\[13pt]
                \sigma_1 & \displaystyle  = \frac{\sqrt{p_2}}{\sqrt{p_1} + \sqrt{p_2}} \\[13pt]
	  \sigma_2 & \displaystyle = \frac{\sqrt{p_1}}{\sqrt{p_1} + \sqrt{p_2}}\\
	\end{array}
	\label{MAXRESULT1}
\end{equation}
	
These relations show the expected invariance under the interchange of
$p_1$ and $p_2$.  To find the boundary limited currents in the simple mean field
approximation we assume that at least one of $\alpha$ or $\beta$ is less than the smaller
of $p_1$ and $p_2$.  For the sake of simplicity, we also assume the lattice is composed
of a whole number of periods, so that the last lattice site is a $p_2$ site.
Focusing initially on the entry-limited case and applying the current
continuity conditions we find, for arbitrary $k$

\begin{equation}
  J = \alpha (1 - \sigma_1) = p_1 \sigma_1 (1 - \sigma_{2}) = \ldots = p_2 \sigma_{k+1} (1 - \sigma_{1})
  \label{SMFT2}
\end{equation}

\noindent Yielding the following densities and currents for the $k = 1$ case:

\begin{equation}
  \begin{array}{ccccccc}
    J_{\alpha}& =& \displaystyle \alpha (1-\sigma_1) & = & p_1 \sigma_1 (1-\sigma_2) & = & \displaystyle p_2 \sigma_2 (1-\sigma_1) \\
    &\Rightarrow&  J_{\alpha} & = & \displaystyle \frac{p_1 (p_2 -\alpha) \alpha}{p_2 (p_1 + \alpha) - \alpha p_1}& = & \alpha(1-\sigma_1)\\
     &          &  \sigma_1 & = & \displaystyle \frac{\alpha p_2}{p_2 (p_1 + \alpha) - \alpha p_1} & &\\                                  
     &          &  \sigma_2 & = & \displaystyle {\alpha}/{p_2} & &
  \label{LEFTRESULT1}
  \end{array}
\end{equation}

\noindent Similarly in the exit limited case we have

\begin{eqnarray}
  J & = & p_1 \sigma_{N - T} (1 - \sigma_{N-T+1}) = p_1 \sigma_{N - T + 1}(1 - \sigma_{N-T+2}) \nonumber \\
    & = &  \ldots = p_1 \sigma_{N-1} (1 - \sigma_N) = \beta \sigma_N \nonumber \\
  \label{SMFT3}
\end{eqnarray}

\begin{equation}
  \begin{array}{cccccccc}
    J_{\beta}&=& \beta (\sigma_2) & = & p_1 \sigma_1 (1-\sigma_2) & = & p_2 \sigma_2 (1-\sigma_1) \\
    &\Rightarrow & J_{\beta} & = & \displaystyle \frac{p_1 (p_2 -\beta) \beta }{p_2 (p_1 + \beta) - \beta p_1} & &\\
    &            & \sigma_1 & = & \displaystyle 1 - \frac{\beta}{p_2} & &\\
    &            & \sigma_2 & = & \displaystyle \frac{p_1(p_2 - \beta)}{p_2 (p_1 + \beta) - p_1 \beta} & &
  \end{array}
  \label{RIGHTRESULT1}
\end{equation}

To determine the transitions between the maximal current and entry limited regions we
equate the maximal current solutions generated by (\ref{SMFT1}) to the expressions for
$J_{\alpha}$ and $J_{\beta}$ in (\ref{SMFT2}) and (\ref{SMFT3})
to find the boundaries between the entry/exit
limited and maximal current phases $(\alpha^{*},\beta^{*})$.  For the $k=1$ case this yields

\begin{equation}
  \begin{array}{ccc}
    \alpha^{*} & = \beta^{*} &  \displaystyle= \frac{\sqrt{p_1} p_2}{\sqrt{p_1} + \sqrt{p_2}} 
  \end{array}
\end{equation}

Finally equating the the current expressions in equations (\ref{SMFT2}) and (\ref{SMFT3}) we find that
the transition between the entry and exit limited regions occurs when $\alpha = \beta$.
    
\subsection{Refined Mean Field Methods}

While appealingly transparent, we will see in the next section that the results of 
the simple mean field approach do
not provide a particularly good match to the results of Monte Carlo simulations.  

We attribute the poor performance of the simple mean field method to the method's
failure to capture the  correlations between the occupation probabilities of different lattice
sites.  To address this deficiency we apply two alternative mean field approaches.
The first refined mean field, which we call the {\it finite-segment} method (FSMFT), involves exactly solving
the master equation for a finite segment of the dual-rate TASEP lattice in a self-consistent manner
(see Appendix \ref{SegmentSection}).  While primarily numerical in nature, this method can
produce estimates for the current and density comparable to the results from MC simulations.
The quality of the estimates produced by the FSMFT approach is a function of the length of the finite-segment,
with longer segments producing superior results.

The second enhanced mean field method is an extension to the two hopping rate
TASEP of the well-known cluster mean field approach described in
\cite{SCHRECKENBERG} and \cite{CHOWDHURY}.  This method tries to
capture the occupancy correlation between adjacent sites by solving the exact master
equation for the multi-site probabilities $P(\tau_i,\tau_{i+1},\tau_{i+2},\ldots)$.  As before $\tau_i$ refers to the
instantaneous occupancy of site $i$ (see Appendix \ref{ClusterSection}).  As is typically the case, the cluster
mean-field method becomes increasingly unwieldy as the size of the cluster increases.  The
results for the $k=1$ case are

\begin{equation}
  \begin{array}{l}
    \sigma_1 = \displaystyle \frac{p_1 + 2p_2}{3(p_1 + p_2)},\,\,
\sigma_2 =  \displaystyle \frac{2p_1 + p_2}{3(p_1+p_2)}, \,\, J  = \displaystyle \frac{p_1 p_2}{2(p_1 + p_2)}
  \end{array}
  \label{CLUSTER6}
\end{equation}

As expected, the solutions are invariant under the exchange $(\sigma_1,p_1) \leftrightarrow
(\sigma_2,p_2)$, and regenerate the standard TASEP results in the limit $p_1 = p_2$. 
We note that in the maximal current solution, $\sigma_1 + \sigma_2 = 1$ as a consequence of
the particle-hole symmetry in the model.

Finally, note that we can use
these results to predict the location of the phase boundary between the maximal
current and entry limited phases.  Using the results for the densities in the maximal
current phase and applying current continuity at the entrance of the lattice we find

\begin{equation}
  \begin{array}{cccc}
               & \displaystyle \alpha^{*} 
    (1 - \frac{p_1 + 2p_2}{3(p_1+p_2)}) & = & \displaystyle \frac{p_1 p_2}{2(p_1 + p_2)}\\[13pt]
   \Rightarrow & \alpha ^{*} = \displaystyle \frac{3p_1 p_2}{2(2p_1+p_2)} &\:  & \\
               &                                                          &    &
  \end{array}
  \label{CLUSTER7}
\end{equation}

\noindent To find the critical value of the extraction rate $\beta^{*}$,
we enforce current continuity at the exit of the lattice

\begin{equation}
  \begin{array}{cccc}
               & \displaystyle \beta^{*} \frac{2p_1 + p_2}{3(p_1+p_2)} & 
          = & \displaystyle \frac{p_1 p_2}{2(p_1 + p_2)}\\[13pt]
   \Rightarrow & \beta^{*} = \displaystyle \frac{3p_1 p_2}{2(2p_1+p_2)} &\:  &
  \end{array}
  \label{CLUSTER8}
\end{equation}
\\
\noindent The fact that $\alpha^{*} = \beta^{*}$ is a reflection of the particle-hole symmetric intrinsic to the 
two-rate TASEP model.

A similar approach (see Appendix \ref{ClusterSection})
can used to find current and density approximations in the parameter
regimes where the entry or exit rate limits the rate of particle transport through the lattice.
First addressing the entry limited region we find

\begin{widetext}
  \begin{equation}
  \begin{array}{ccl}
    J_{\alpha} & = &  \displaystyle \frac{\alpha(p_1 (p_2 - \alpha) 
   -\alpha^2 + \sqrt{4{p_1^2}(p_2 - \alpha)\alpha + (p_1p_2 - p_1\alpha + \alpha^2)^2})}{2p_2 (\alpha + p_1)} \\[13pt]
    \sigma_1 & = & \displaystyle  \frac{2\alpha p_2 + \alpha^2 + {p_1}(\alpha + {p_2}) 
        - \sqrt{4{p_1^2}(p_2 - \alpha)\alpha + (p_1p_2 - p_1\alpha + \alpha^2)^2}}{2p_2 (\alpha + p_1)} \\[13pt]
    \sigma_2 & = & \displaystyle \frac{\alpha}{p_2}
  \end{array}
  \label{ALPHA2}  
\end{equation}

\noindent Similarly, the solution in the exit-limited phase is

\begin{equation}
  \begin{array}{ccl}
    J_{\alpha} & = &    \displaystyle \frac{\beta(p_1 (p_2 - \beta) 
    - \beta^2 + \sqrt{(4{p_1^2}(p_2 - \beta)\beta + (p_1p_2 - p_1\beta + \beta^2)^2})}{2p_2 (\beta + p_1)} \\[13pt]
    \sigma_1 & = & \displaystyle 1 - \frac{\beta}{p_2} \\[13pt]
    \sigma_2 & = &  \displaystyle \frac{p_1 (p_2 - \beta) 
         - \beta^2 + \sqrt{4{p_1^2}(p_2 - \beta)\beta + (p_1p_2 - p_1\beta + \beta^2)^2}}{2p_2 (\beta + p_1)}
  \end{array}
  \label{BETA2}  
\end{equation}
\end{widetext}
  
\section{Monte Carlo Simulations}

Extensive Monte Carlo simulation were performed to validate the various analytical models presented in the
previous section.  As we expected the densities in the lattice to vary significantly as the internal hopping rates were
varied, we decided to base our Monte Carlo code on the BKL continuous-time algorithm \cite{BORTZ75}.  The BKL
algorithm has the advantage of maintaining a constant computational efficiency over a wide range of particle
densities.

The magnitude of the finite size effect in our simulations was estimated by running lattices 
of varying lengths.  For lattices at least one thousand sites long, the MC
results were found to systematically deviate from the known TASEP results ($p_1 = p_2$)
by less than half a percent.  As a result, we used lattices containing approximately 1000 sites for
all our simulations. The simulations were run for $4 \times 10^9$ steps, which was sufficient to
reproduce the known TASEP results in lattices much longer than our 1000 site
standard.  In all simulations, $p_1$ was normalized to $1$.
Finally, to ensure an unbiased sampling of the lattice states, a linear-congruential 
pseudorandom number generator with a period of $2 \times 10^{18}$ was used \cite{NR}.

\subsection{Currents}
\label{CURRENTSECTION}

\begin{figure}[h]
  \includegraphics[width=2.9in]{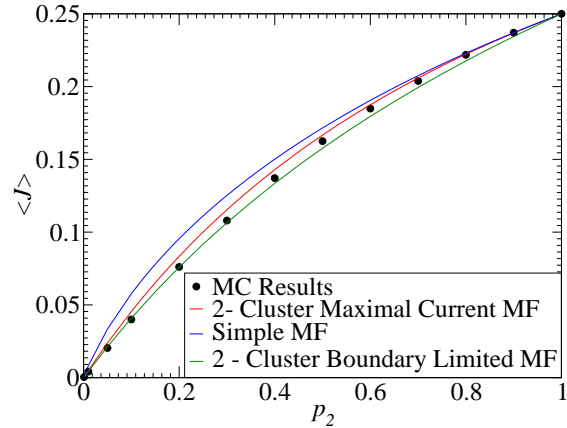}
  \caption{Current in the maximal current phase for a $T=2$ dual-rate TASEP.
           The results from the 2-cluster boundary limited MFT were produced
           by maximizing $J$ in (\ref{ALPHA2}) with respect to $\alpha$.}
  \label{CURRENTS1}
\end{figure}

\begin{figure}[h]
  \includegraphics[width=2.9in]{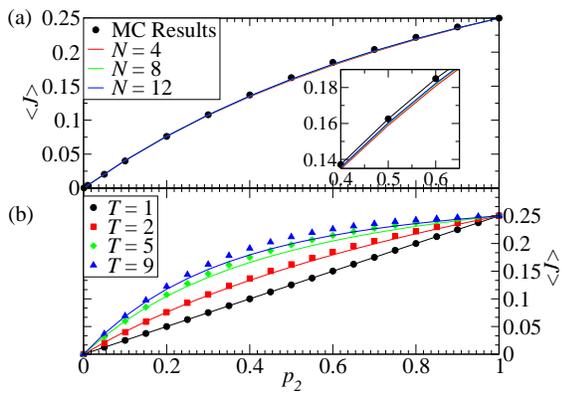}
  \caption{(a) The effect of increasing segment length on the quality of the FSMFT estimates
           of the maximal current in a $T=2$ dual-rate TASEP.  The points are the Monte-Carlo
           simulation results, the lines are the FSMFT estimates.  For segment lengths greater
           than or equal to twice the period, the quality of the current estimate shows only
           marginal improvement with increasing length. 
           (b) Maximal currents for dual-rate TASEPs of various periods.  The FSMFT estimates
           were produced using a single period of the lattice as the finite segment.  While
           providing reasonable estimates, the deviation between the MC and FSMFT results
           increases as the period ($T$) increases.  As seen in (a), this increase in error
           can be mitigated by increasing the number of periods included in the finite-segment.}
  \label{CURRENTS2}
\end{figure}

The maximal current results (Figs \ref{CURRENTS1}, \ref{CURRENTS2}) 
show the expected qualitative behavior with
a low current for small values of $p_2$ and a value of $1/4$ when $p_2 = p_1 = 1$. 
In Fig \ref{CURRENTS1} the boundary-limited current predictions (Eqn
\ref{ALPHA2}) were used to generate a prediction for the maximal current as a
function of $p_2$.  This was accomplished by maximizing $J$ in Eqn
\ref{ALPHA2} with respect to $\alpha$.  Consistent with our expectations, we find
that the two cluster MF approaches, and the finite-segment mean field approach,
provide better approximations to the Monte-Carlo results than does the simplest
mean-field method.  The relatively poor performance of
the simple mean field model can be ascribed to the strong pair correlations present
in the two-rate TASEP, which the simple MF model completely fails to capture.  
The 2-cluster maximal current mean field clearly
performs best when $p_{2}/p_{1} \approx 1$.  In this, limit the model is
essentially the normal single rate TASEP, and the expected pair correlations
between adjacent $p_1$ and $p_2$ sites are negligible.  As simple mean field
methods are known to provide exact results for the standard TASEP, the ability of
the 2-cluster MF to provide accurate results in this region is not surprising.  In
contrast the boundary-limited 2-cluster MF performs best when $p_2 \ll p_1$
where the correlations between adjacent $p_1$ and $p_2$ sites are expected to
be strongest (although the known TASEP results are returned with $p_1$ = $p_2$).  
Nonetheless in the limit $p_2 \rightarrow 0$, the {\it relative} error increases for all the 
mean field methods.

\begin{figure}[h]
  \includegraphics[width=2.9in]{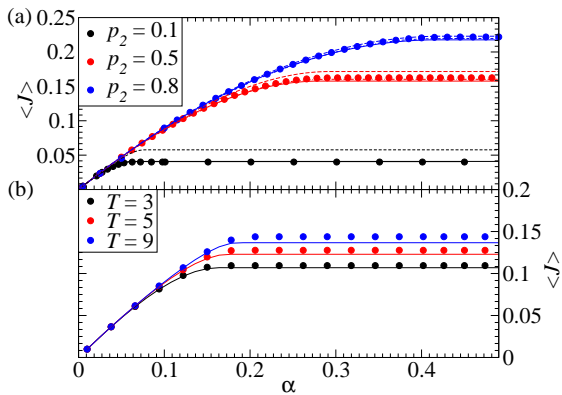}
  \caption{(a) Current profiles along the $\alpha$ direction in the $T=2$ dual-rate TASEP phase plane.
           The solid lines were produced using (\ref{ALPHA2}), the dashed lines were produced
           using (\ref{LEFTRESULT1}), and the dotted-dashed lines were produced using an $N=8$
           FSMFT approach.
           (b) Current profiles for dual-rate TASEPS of various periods with $p_2 = 0.25$.
           The FSMFT results (solid lines)
           were produced using an $N=T$ FSMFT approach.  From the plot we can observe the second order
           character of the entry-limited to maximal current transition.}
  \label{ALPHACURRENT1}
\end{figure}

\subsection{Densities}
\label{DENSITYSECTION}

\begin{figure}[h]
  \includegraphics[width=2.9in]{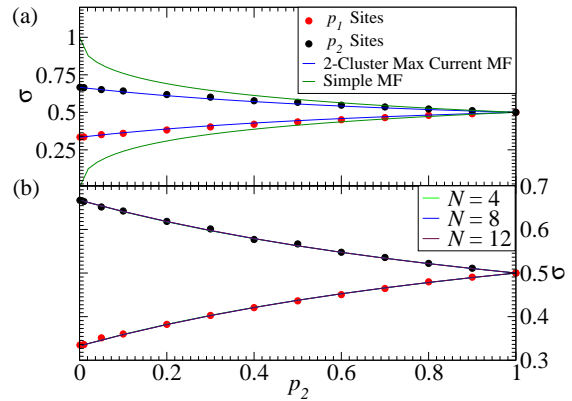}
  \caption{(a) Densities from the simple MF method, the 2-cluster mean field method, and the Monte Carlo (MC)
   simulations for a $T=2$ dual-rate TASEP.  The simple mean field assumption results in the largest error
   for small values of $p_2$ where we expect the density correlation between adjacent sites to be large.
   (b)  Results produced by the FSMFT method for various segment sizes.}
  \label{DENSITY1}
\end{figure}

\begin{figure}[h]
  \includegraphics[width=2.9in]{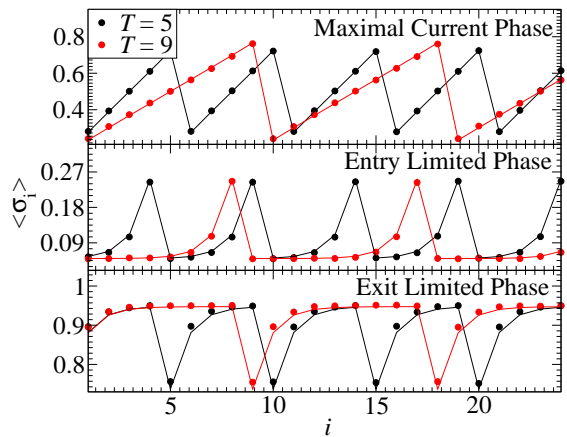}
  \caption{The bulk density profiles for $T=5$ and $T=9$ dual-rate TASEPS with $p_2/p_1$ = 0.2,
   in all three current phases.  The plots show the good agreement between the Monte-Carlo (points)
   and $N=T$ FSMFT (solid lines) results.}
  \label{DENSITY2}
\end{figure}

\begin{figure}[h]
  \includegraphics[width=2.9in]{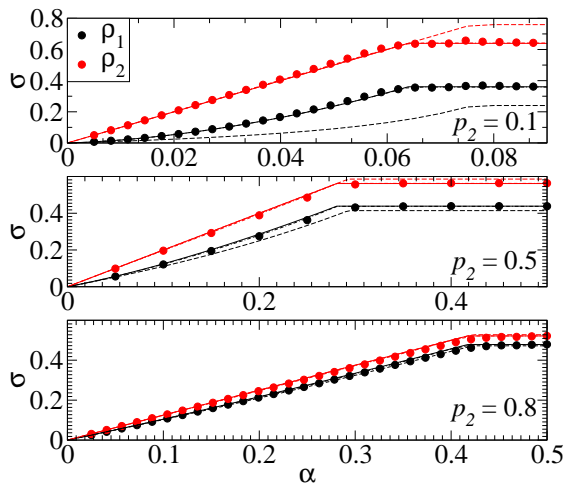}
  \caption{The density profile along the $\alpha$ direction in the $T=2$ dual-rate TASEP phase plane.  The solid
  lines were produced using (\ref{ALPHA2}), the dashed lines were produced using 
  (\ref{LEFTRESULT1}), while the dotted-dashed lines were produced by an $N = 8$ finite-segment MF approach.
  Note that although solution for $\sigma_2$ is the same in Eqns. (\ref{LEFTRESULT1}) and
  (\ref{ALPHA2}), the predicted value of $\alpha^{*}$ differs between the two mean field theories.
  This is the reason for the significant difference between the $\sigma_2$ profiles predicted by the two
  mean field methods at small values of $p_2$.}
  \label{ALPHADENSITY}
\end{figure}

Referring to Figs \ref{DENSITY1} through \ref{ALPHADENSITY},
we find the finite-segment MF, and both 2-cluster mean field
methods provide excellent matches to the simulation results.
The densities produced by the maximal current 2-cluster MFT, the boundary limited 2-cluster MFT, and the finite
segment method are all within 5\% of the MC results, with the FSMFT results improving with increased segment length. 
For all three approaches the quality of the agreement is relatively uniform over all the values of $p_2$,
though the maximal current
2-cluster mean field results do tend to have an increased relative error at smaller values of $p_2$ where the
correlations between adjacent sites are expected to be strongest. In contrast,
the simple mean field approach shows very
significant deviations from the experimental results particularly the small $p_2$
limit.  Given the simple mean field method's inability to capture correlations
in the density between adjacent fast and slow sites, the poor performance of this
method at small values of $p_2$ is not unexpected. 

\subsection{Correlations}
\label{CORRELATIONSECTION}

\begin{figure}[h]
  \includegraphics[width=2.9in]{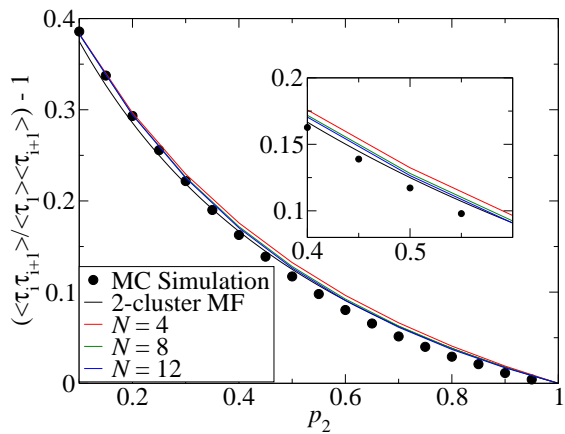}
  \caption{The pair correlation function produced by the MC simulations and the various MF approximations
  for the $T=2$ dual-rate TASEP.
  As expected the 2-cluster mean field approach provides a very good match to the MC results.  
  As expected, the results produced by the segment MF improve with an increasing segment length.}
  \label{CORRELATION1}
\end{figure}

\begin{figure}[h]
  \includegraphics[width=2.9in]{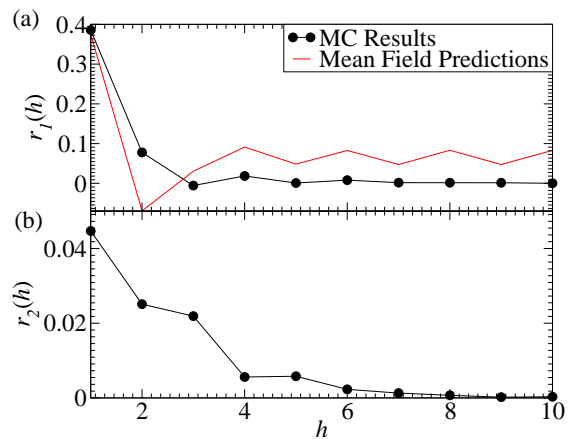}
  \caption{The pair correlation function produced by the MC simulations for the $T=2$ dual-rate TASEP.
  Plot (a) displays the
  correlation between fast sites and their neighbors to the right, while (b) does the same
  for slow sites. In both plots $p_2 = 0.1$
  In (a), the pair correlation ($h = 1$) is quite strong indicating that the slow sites are acting to limit
  the rate of transport through the lattice consistent with the maximal current assumption.  Conversely, (b)
  shows that slow sites are essentially uncorrelated with their neighbors to the right.}
  \label{CORRELATION2}
\end{figure}

Looking at the maximal-current density correlations displayed in Figs
\ref{CORRELATION1} and \ref{CORRELATION2}, we see that the pair
correlation $\langle \tau_1 \tau_2\rangle $ is very large for small values of $p_2$
and approaches zero as $p_2 \rightarrow p_1$.  The size of this pair correlation for
small $p_2$ confirms our intuition that the slow sites dominate the dynamics of the
system, and is consistent with the maximal current assumption.

\begin{figure}[h]
  \includegraphics[width=2.9in]{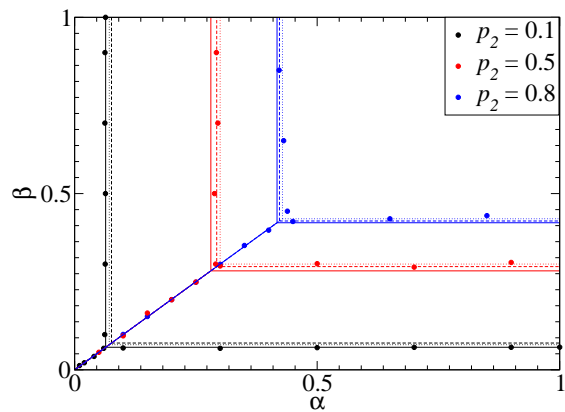}
  \caption{The phase diagram for the $T=2$ two-rate TASEP.  The dashed lines are produced by the simple mean field
  theory (Eqns \ref{LEFTRESULT1}, \ref{RIGHTRESULT1}).  The dotted lines are produced by the maximal
  current 2-cluster method (Eqn \ref{CLUSTER6}), and the solid lines are produced by the boundary-limited
  mean field method (Eqns \ref{ALPHA2}, \ref{BETA2}).}.
  \label{PHASEDIAGRAM}
\end{figure}

Figure \ref{CORRELATION2}(b) indicates that in the maximal current
phase the $\langle \tau_2 \tau_1\rangle$ pair correlation is negligible.  
The absence of significant correlation in Fig
\ref{CORRELATION2}(b) and the rapid decay of the correlation function in Fig
\ref{CORRELATION2}(a) indicates that the dynamics of the two-rate TASEP are
essentially the dynamics of a set of successive, and nearly independent
$(p_1,p_2)$ unit cells.  While the correlation functions in Fig.
\ref{CORRELATION2} show longer range behavior, sites separated by
more than two intervening lattice sites are essentially uncorrelated.
Despite the maximal current 2-cluster method's ability to provide highly accurate estimates for the pair
correlation function (Fig. \ref{CORRELATION1}), the 2-cluster method generates predictions for the
current and density that carry small but significant error particularly for small $p_2$.
To seek the source of this error, we must consider the correlation profile produced by the maximal-current 
2-cluster method.  If we consider two lattice sites separated by $h$ intervening lattice sites,
the general pair correlation function can then be formed as follows:

\begin{equation}
  \begin{array}{ccc}
    r(h) & = & \langle \tau_i \tau_{i+h}\rangle /\langle \tau_i\rangle \langle \tau_{i+1}\rangle - 1\\
    & & \\
    \mathrm{Where:} & & \\
    \langle \tau_{i} \tau_{i+h}\rangle & = & \displaystyle \frac{\sum_{\tau_i} P(1,\tau_1)\left[\Pi_{i=2}^{h-2} P(\tau_i,\tau_{i+1})\right]P(\tau_{h-1},1)}{\sum_{\{\tau_i\}} \Pi_{i=1}^{h-1} P(\tau_i,\tau_{i+1})}
  \end{array}
\end{equation}

Referring to Fig. \ref{CORRELATION2}, we see that while the 2-cluster method provides good estimates for the pair
correlation function, its predictions for the correlation profile are relatively poor.  While the correlation profile
produced from the results of the MC simulations decays rapidly to zero, the results are not negligible for $h = 2, 3$,
and thus we expect these small longer range correlations to have some effect on the current and density of the two site
TASEP.  The failure of the maximal current 2-cluster method to capture these weak longer range
correlations may be responsible for the increasing error in the mean field current and densities with decreasing $p_2$.

\subsection{Phase Diagram}

The phase diagram derived from simulations is displayed in Fig \ref{PHASEDIAGRAM}, along with the phase
boundaries predicted by the various mean field methods.  The Monte Carlo phase boundaries were computed
by taking numerical derivatives of the current as a function of $\alpha$ and $\beta$ and locating any
clear discontinuities in the derivatives.  The dual-rate TASEP retains the general
form of the standard TASEP phase-diagram, and the order of the transitions remains unchanged; first order
between the high and low density regimes, and second order between the boundary limited and maximal current
regions.  The most significant deviation from the standard phase diagram is the
increase in the maximal current region which accompanies a decrease in one of the hopping rates ($p_2$ in our example).
Physically, the maximal current region is defined as the region of the $(\alpha,\beta)$ parameter space where the
internal motions determine the net particle flow through the lattice.  The increase in the area of the maximal current
phase with decreasing $p_2$ is then expected as we naturally require $\alpha$ or $\beta$ to be less than $p_2$ in order
to enter a boundary limited region.

Despite the varying degrees of success in predicting accurate steady state currents and densities, all
three mean field approaches predict the phase boundaries within approximately $12\%$ error.  As was observed in
the mean field predictions for the steady state currents and densities, the boundary-limited 2-cluster mean field
approach provides the best predictions at small values of $p_2/p_1$, while the maximal current mean field method
excels when $p_2 \approx p_1$.  We also note that the phase transitions predicted by the boundary-limited 2-cluster
approach were determined numerically by maximizing $J$ in Eqns (\ref{ALPHA2}) and (\ref{BETA2})

\section{Conclusion}

We have developed and presented four separate approximate methods to compute the
the current and densities of totally asymmetric exclusions processes involving two
internal hopping rates.  Moreover, we have extensively simulated the
two-rate TASEP and have thoroughly explored its phase diagram.  
We find that the dual-rate TASEP retains the three phases
found in the standard TASEP model.  Within each of these phases the best of our
mean field theories provide highly accurate, albeit not exact, predictions for the
particle currents and densities.  In particular, a maximal current phase 2-cluster
mean field theory was developed that provides accurate estimates (Eqn.
\ref{CLUSTER6}) for the current and the density (Figs.
\ref{CURRENTS1},\ref{DENSITY1}) and the pair correlation function $\langle
\tau_1 \tau_1\rangle $ (Fig. \ref{CORRELATION1}).  Similarly, in the boundary
limited phases a simplified 2-cluster MF method was introduced (Eqn.
\ref{ALPHA2}, \ref{BETA2}), which produced accurate estimates for the boundary
currents, and reasonably accurate estimates for the maximal currents (Fig.
\ref{ALPHACURRENT1}).  In addition to the generally
inferior simple mean field approach (Eqns. \ref{MAXRESULT1},
\ref{LEFTRESULT1}, \ref{RIGHTRESULT1}), we have developed a primarily
numerical mean field method that we have termed the finite segment mean field
method.  This method can rapidly produce accurate estimates for the currents
and densities in all three phases of the dual-rate TASEP.  Additionally, the
finite-segment method shows promise as a means of producing accurate estimates for the
current and the density of TASEPs with hopping rates that vary spatially
in a manner more complex than the periodic pattern described in this paper.

Given the success of mean field theories in predicting the current and density
in the standard TASEP, as well as various extended TASEPs, the inability of our
mean field approaches to provide exact results is noteworthy.  
Comparison of our best mean field predictions of the density correlation function 
with the Monte Carlo results (Figs \ref{CORRELATION1},
\ref{CORRELATION2}) indicate that while the 2-cluster mean field approach
accurately captures the strong correlation between adjacent fast and slow
lattice sites, the method does not account for the weaker, but still
significant, longer range correlations.  An extension to a three or four site
cluster model may yield better analytic results.

\begin{acknowledgments}
GL acknowledges support from the Natural Science and Engineering Research Council of Canada.
TC acknowledges support from the NSF (DMS-0206733), and the NIH (R01 AI41935). 
ABK acknowledges support from the Camille and Henry Dreyfus New Faculty
Awards Program (under Grant No. NF-00-056), from the Welch
Foundation (under Grant No. C-1559), and from the US National Science
Foundation through the grant CHE-0237105. The authors thank B. Bergersen, 
and G. Patey for valuable comments on the
manuscript and C. Greif for his assistance with the Arnoldi diagonalization routines.
\end{acknowledgments}

\appendix

\section{The Finite-Segment Mean Field Method}
\label{SegmentSection}

Here we give a detailed description of the the algorithm underlying the
finite segment mean field method, and focus on using the method to solve for the currents
and densities in the maximal current phase.  Consider a finite lattice containing
$N$ sites and a whole number of $(p_1,p_1,\ldots,p_2)$ periods.  Working
under the assumption that the bulk densities in the lattice are periodic and inherit the
same periodicity as the movement rates, we focus on a single period.
We then define the effective rate of injection
into this finite segment of the lattice and the effective rate of extraction in terms of the unknown particle
densities inside the segment.  This yields $\alpha_{\mathrm{eff}} = p_2 \sigma_{k+1}$ and
$\beta_{\mathrm{eff}} = p_2 (1 - \sigma_1)$.  Using these rates and the
known values of $p_1$ and $p_2$ we can build the transition matrix for the master equation
describing the motion of the particles in our $T$-site finite-segment.  Letting $N$ be the number
of sites in the finite segment, and $\vec{R}_i$ be the movement rate for site $i$ in the segment,
the algorithm to construct the master equation transition matrix is (Fig. \ref{ALGORITHM1}):

\begin{figure}[h]
  \includegraphics[width=2.9in]{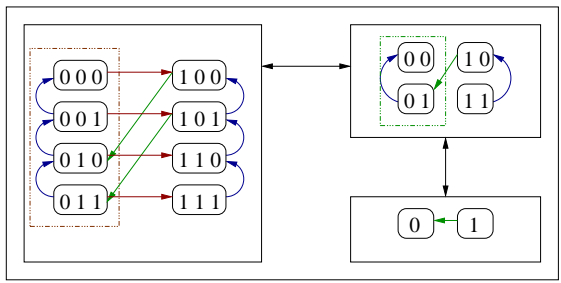}
  \caption{The steps in the algorithm to generate the transition matrix for a TASEP model.
     For the purposes of illustration a three site model has been used.  To begin we treat each
     possible occupancy of the lattice as a bit pattern, and label the state with the corresponding
     decimal value.  For example given that \texttt{011} is the binary representation of \texttt{3}, we label the
     state with particles occupying the second and third lattice sites as \textit{state 3}.  Next we
     divide the states into two groups; states where the
     first lattice site is occupied \textit{(1-states)}, and states where the first lattice site is empty 
     \textit{(0-states)}.
     Regardless of the number of lattice sites in the TASEP, the transitions between the two classes of 
     states \textit{always} occur between the first half of the \textit{1-states} 
     and the second half of the \textit{0-states}
     (the straight green arrows in the figure).
     To determine the remaining transitions we call the algorithm recursively
     on both the \textit{1-states} and
     the \textit{0-states}, making sure to ignore the highest order bit ({\it i.e.} the leftmost lattice site).  Finally
     we add in the trivial transitions between each 
     \textit{0-state} and each \textit{1-state} that are the result of injection
     at the left edge of the lattice (the red arrows in the diagram).  With the connectivity of the
     states fully enumerated, assigning the appropriate rate $p_i$ to each of the transitions is straightforward.}
  \label{ALGORITHM1}
\end{figure}

\begin{enumerate}

\item Label every state of the $N$-site finite segment with a number determined by the
occupancy of the lattice sites.  For example, an $N = 3$ lattice with only the last
site filled ({\it ie} an occupancy of $001$), is said to be in {\it state 1}, while the
same lattice with the last two sites filled (an occupancy of $011$) is said to be in
{\it state 3}. \label{STEP1}

\item Separate the states generated in step \ref{STEP1} into two groups; the states where
the left-most lattice site is empty (the {\it 0-states}), and the state where the left-most
lattice site is full (the {\it 1-states}). \label{STEP2}

\item Enumerate the transitions between the 1-states and the 0-states.  The transitions will
always be the movements that take the system from the lowest $2^{N-2}$ 1-states to the highest
$2^{N-2}$ 0-states.  The rate of these transitions is $\vec{R}_l$, where $l$
is the current depth of the recursion (see step \ref{STEP4}). \label{STEP3}

\item Recurse on the 0-states and then on the 1-states, ignoring the leftmost site ({\it ie} with each
recursion of the algorithm, we reduce the length of the segment by 1), or stop if the current
segment length is $N=1$.\label{STEP4}

\item Once the recursion specified in steps \ref{STEP1} through \ref{STEP4} has completed,
enumerate all the trivial transitions between the 0-states and the 1-states that occur at rate $\alpha$
({\it ie} the state transitions due to particle injection on the left).

\end{enumerate}

With the master equation constructed by this algorithm, we can
then compute the densities in the finite segment, apply a self consistency condition on the densities,
and subsequently compute the current in terms of the hopping rates $p_1$ and $p_2$.  Thus utilizing the
state enumeration scheme and algorithm described in Fig. \ref{ALGORITHM1}, we generate and solve the
following equations

\begin{equation}
  \begin{array}{lcc}
     {\bf M(\{\sigma_i\},p_1,p_2)}\vec{V}& = & 0\\
     \sum_{i={2^{N-1} + 1}}^{2^N} V_i(\{\sigma_i\},p_1,p_2) & = &\sigma_1 \\
     \sum_{i=1}^{2^{N-1}} V_{2i-1}(\{\sigma_i\},p_1,p_2) & = &\sigma_N \\
  \end{array}
  \label{FIXEDPOINT1}
\end{equation}
      
In Eqn (\ref{FIXEDPOINT1}), $\bf{M}$ is the transition matrix, while $V_i$ is the probability of finding the
finite-segment in occupancy state $0 < i < (2^N - 1)$ (see Fig. \ref{ALGORITHM1}).  
When applying the finite-segment mean
field approach, (\ref{FIXEDPOINT1}) is solved iteratively until a fixed point in the densities is 
reached.  Using the results from (\ref{FIXEDPOINT1}) we can compute the current
from the expression $J=\alpha_{\mathrm{eff}} \sum_{i=1}^{2^{N-1}} V_{i}$ 
(or $\beta_{\mathrm{eff}}\sum_{i=1}^{2^{N-1}} V_{2i - 1}$).
Additionally, the finite-segment method can be extended to treat the boundary limited dual-rate TASEP by fixing
$\alpha_{\mathrm{eff}} = \alpha$ or $\beta_{\mathrm{eff}} = \beta$ as appropriate,
and applying the density self-consistency
condition at the end of the finite-segment that lies in the interior of the TASEP lattice.
While a simple and effective approach, the exponential increase in the size of the transition matrix with the length
of the segment quickly renders (\ref{FIXEDPOINT1}) analytically untreatable.  As a result we chose to solve
(\ref{FIXEDPOINT1}) numerically using a simple fixed-point iteration scheme.  Employing the fast Arnoldi-method
eigensolvers in the well-known ARPACK linear-algebra software library \cite{ARPACK}, we could
easily treat segments containing $\sim 20$ lattice sites.

\section{Cluster Mean Field Methods}
\label{ClusterSection}

\begin{figure}[h]
  \begin{center}
    \includegraphics[width=2.55in]{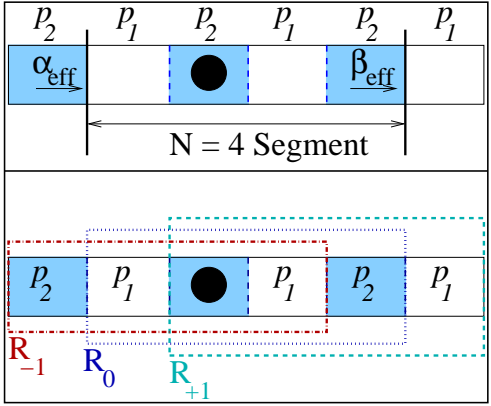}
  \end{center}
\vspace{-6mm}
  \caption{The lattice site groupings used in the two enhanced mean field approaches.  (a)
  The arrangement used for an $N = 4$ finite segment mean field (MFT) approach.  The master
  equation for the state of the four marked lattice sites is solved exactly.  The
  four sites are coupled to the rest of the lattice by assuming an effective injection rate of 
  $\alpha_{\mathrm{eff}} = p_2 \sigma_2$ on the left, and an 
  effective extraction rate of $\beta_{\mathrm{eff}} = p_2 (1 - \sigma_1)$ on the right.
  (b) The
  arrangement of lattice sites used in the 2-cluster mean field theory.  The two lattice sites
  of interest are the central two sites in grouping $R_0$.  Groupings $R_{+1}$ and $R_{-1}$ are
  produced by shifting the boundaries of the group $R_0$ one lattice site to the right, and one
  lattice site to the left respectively.  To generate a second set of equations, we simply interchange
  all the $p_1$ and $p_2$ sites, while keeping the groupings $R_{+1}$, $R_0$, and $R_{-1}$ the same.}
  \label{2PSETS}
\end{figure}

We begin the derivation of the current and density approximations displayed in Eqn
(\ref{CLUSTER6}) by considering the two site probability $P(\tau_i,\tau_{i+1})$.
Here, $\tau_i$ refers to the occupancy of site $i$,
and is equal to 1 if the site is occupied, 0 if the site is empty.  The time evolution of the
occupancy state of any two adjacent sites in a TASEP will depend on the two sites
themselves along with the site immediately to the left and the right of the two site
grouping (Fig \ref{2PSETS}b).

\noindent Thus the master equation for the two site probability is
\cite{CHOWDHURY}

\begin{equation}
     \begin{array}{l}
       \displaystyle  \frac{dP(\tau_i,\tau_{i+1})}{dt}  = \\
         \\
         \sum_{\tau_{i-1},\tau_{i+2}} Q(\tau_{i-1},\tau_{i+1},\tau_i,\tau_{i+2})(p_1\delta_{\tau_i,0}\delta_{\tau_{i+1},1})\\
         - Q(\tau_{i-1},\tau_i,\tau_{i+1},\tau_{i+2})(p_1\delta_{\tau_i,1}\delta_{\tau_{i+1},0})\\[13pt]
         + \sum_{\tau_{i-1},\tau_{i-2}} P(\tau_{i-2},\tau_i,\tau_{i-1},\tau_{i+1})(p_2\delta_{\tau_i,1}\delta_{\tau_{i-1},0})\\
         - P(\tau_{i-2},\tau_{i-1},\tau_i,\tau_{i+1})(p_2\delta_{\tau_{i-1},1}\delta_{\tau_i,0})\\[13pt]
         + \sum_{\tau_{i+2},\tau_{i+3}} P(\tau_i,\tau_{i+2},\tau_{i+1},\tau_{i+3})(p_2\delta_{tau_{i+1},0}\delta_{\tau_{i+2},1})\\
         - P(\tau_i,\tau_{i+1},\tau_{i+2},\tau_{i+3})(p_1\delta_{\tau_{i+1},1}\delta_{\tau_{i+2},0})\\[13pt]
     \end{array}   
     \label{CLUSTER1}
  \end{equation}

As we have incorporated two different
hopping rates into the TASEP model we are forced to deal with two sets of probabilities;
the first $(P(\tau_i,\tau_{i+1},\tau_{i+2},\tau_{i+3}))$
refers to the probability of having configuration $(\tau_i,\tau_{i+1},\ldots)$ given that the first site (site $i$)
is a $p_1$ site. The second probability term $(Q(\tau_i,\tau_{i+1},\tau_{i+2},\tau_{i+3}))$, 
is the probability of configuration $(\tau_i,\tau_{i+1},\ldots)$ when the
first site is a $p_2$ site.  Now we assume we can factor the $4$-site probabilities as follows

\begin{equation}
  \begin{array}{lll}
    P(\tau_i,\tau_{i+1},\tau_{i+2},\tau_{i+3}) & = & P(\tau_i,\tau_{i+1})Q(\tau_{i+1},\tau_{i+2})P(\tau_{i+2},\tau_{i+3})\\
    Q(\tau_i,\tau_{i+2},\tau_{i+2},\tau_{i+3}) & = & Q(\tau_i,\tau_{i+1})P(\tau_{i+1},\tau_{i+2})Q(\tau_{i+2},\tau_{i+3})\\
  \end{array}
\end{equation}

\noindent And by definition we have

\begin{equation}
  \begin{array}{lll}
        \sigma_1 & = & P(1,0) + P(1,1)\\
        \sigma_2 & = & Q(1,0) + Q(1,1)\\
        P(0,0) & = & 1 - P(0,1) - P(1,0) - P(1,1)\\
  \end{array}
  \label{CLUSTER2}
\end{equation}
  
\noindent Thus, in the steady state we can write

\begin{equation}
  \begin{array}{ccl}
    \displaystyle \frac{dP(0,0)}{dt} & = & p_2 \left(P(0,1,0,1)-P(1,1,0,0)\right)\\[13pt]
	  & = & p_2(P(0,1)Q(1,0)P(0,1) \\[13pt]
	  &  \:  & \hspace{5mm}  - P(1,1)Q(1,0)P(0,0)) = 0\\[13pt]\\
	\end{array}
	\label{CLUSTER2A}
\end{equation}

\noindent Upon using (\ref{CLUSTER2}), we find

\begin{equation}
      P(1,0) = \displaystyle \frac{\sigma_1(1-\sigma_2) - (\sigma_2-\sigma_1)^2}{1 + (\sigma_2 - \sigma_1)}
\label{CLUSTER2b}
\end{equation}

\noindent The equation for $Q(1,0)$ can be found from (\ref{CLUSTER2b}), by interchanging $p_1$ and
$p_2$, as well as $\sigma_1$ and $\sigma_2$.  Applying the current continuity condition $p_1 P(1,0) = p_2 Q(1,0)$,
we then arrive at the relation

\begin{equation}
  \begin{array}{ccc}
   \displaystyle p_1 \frac{\sigma_1(1-\sigma_2) - (\sigma_2-\sigma_1)^2}{1 + (\sigma_2 - \sigma_1)} & = & 
   \displaystyle  p_2 \frac{\sigma_2(1-\sigma_1) - (\sigma_1-\sigma_2)^2}{1 + (\sigma_1 - \sigma_2)} \\
  \end{array}
  \label{CLUSTER3}
\end{equation}

\noindent We now make the substitution $\sigma_2-\sigma_1 = k$ and define $p_1$ to be the larger of
the two rates.  Thus $\sigma_1 < \sigma_2$ and $k > 0$.  
Solving (\ref{CLUSTER3}) for $\sigma_1$, we find $ \sigma_1=$

\begin{widetext}
  \begin{equation}
    \begin{array}{l}
  \displaystyle \frac{\left[p_2(1-k^2) - p_1(k-1)^2\right] + \sqrt{(1-k^2)(k(p_1+p_2)- (p_1 - p_2))(3k(p_1+p_2) - (p_1 - p_2))}}
                     {2k(p_1 + p_2) - (p_1 - p_2)}\\
    \end{array}
    \label{CLUSTER4}
  \end{equation}
\end{widetext}

\noindent From (\ref{CLUSTER4}) we find that $\sigma_1$ is real only 
when $k < \frac{p_1-p_2}{3(p_1+p_2)}$ or
$k > \frac{p_1-p_2}{p_1+p_2}$.  Substituting (\ref{CLUSTER4}) into (\ref{CLUSTER3}) we find
$J=\frac{k p_1 p_2}{(p_1 - p_2) - k(p_1+p_2)}$, which shows $J < 0$ for $k > \frac{p_1 -p_2}{p_1+p_2}$,
$J > 0$ for $0 < k < \frac{p_1 - p_2}{p_1 + p_2}$, and 
$\frac{dJ}{dk} = \frac{p_1 p_2 (p_1 - p_2)}{(k(p_2+p_1)+(p_2-p_1))^2} > 0 \forall k$

\noindent Since $J$ must be positive, $\sigma_1$ must be real, and $J$ is monotonically 
increasing, the maximum value of the current
must occur when $k = \frac{p_1-p_2}{3(p_1+p_2)}$.  Substituting this value of $k$ into (\ref{CLUSTER4}),
we arrive at (\ref{CLUSTER6}).

An analogous approach can be used to find current and density estimates in the entry and exit limited regions.
Considering the entry limited case first, we find the following master equation for the 
occupancy of the first two sites in the lattice:

\begin{equation}
  \begin{array}{ll}
 \displaystyle    \frac{dP(0,0)}{dt} = & -\alpha P(0,0) + p_2 P(0,1)(P(0,0) + P(0,1)) \\[13pt]
 \displaystyle     \frac{dP(0,1)}{dt} = & -\alpha P(0,1) - p_2 P(0,1)(P(0,0) + P(0,1)) \\[13pt]
                         & \hspace{1cm} + p_1 P(1,0) \\[13pt]
 \displaystyle     \frac{dP(1,0)}{dt} = & \alpha P(0,0) + p_2 P(1,1)(P(0,0) + P(0,1)) \\[13pt]
                         &  \hspace{1cm}  - p_1 P(1,0) \\[13pt]
   \displaystyle   \frac{dP(1,1)}{dt} = & -p_2 P(1,1)(P(0,0) + P(0,1)) + \alpha P(0,1) \\
  \end{array}
  \label{ALPHA1}
\end{equation}

In writing (\ref{ALPHA1}) we have assumed that the occupancy
probabilities $P(\tau_1,\tau_2)$ do not vary significantly as a function of position near the
entrance of the lattice.  This assumption is analogous to that used with the standard
TASEP in the boundary limited region where the density is assumed to be essentially constant.  
Applying the normalization
condition on the probabilities $P(\tau_i,\tau_{i+1})$ and solving (\ref{ALPHA1}) generates
the results displayed in (\ref{ALPHA2}).  An analogous approach yields
the current and densities in the exit limited regime (Eqn. \ref{BETA2}).

\end{document}